\documentclass[10pt, a4paper]{scrartcl}

\usepackage{amsmath,amssymb}
\usepackage[utf8x]{inputenc}
\usepackage{textcomp,marvosym}
\usepackage{cite}
\usepackage{nameref,hyperref}
\usepackage[right]{lineno}
\usepackage[table]{xcolor}
\usepackage{array}

\usepackage{lastpage,fancyhdr,graphicx}
\usepackage{epstopdf}
\begin{document}

\noindent\hrulefill 
\vspace{2mm}
{\Large
\begin{center}
\textbf{Dynamics of economic unions:\\
an agent-based model to investigate\\the economic and social drivers of withdrawals
}
\end{center}
}
\vspace{2mm}
\noindent\hrulefill 
\vspace{2mm}

\begin{center}
Carlos Gracia-L\'azaro\textsuperscript{1},
Fabio Dercole\textsuperscript{2},
Yamir Moreno\textsuperscript{1,3,4}\\
\bigskip
\textbf{1} Institute for Biocomputation and Physics of Complex Systems, University of Zaragoza, 50018, Zaragoza, Spain.
\\
\textbf{2} Department of Electronics, Information, and Bioengineering, Politecnico di Milano, 20133, Milano, Italy.
\\
\textbf{3} Department of Theoretical Physics. University of Zaragoza, 50005, Zaragoza, Spain.
\\
\textbf{4} ISI Foundation, Turin, Italy.
\\
\bigskip
April 25, 2022
\end{center}



\section*{Abstract}

Economic unions are international agreements oriented to increase economic efficiency and establishing political and cultural ties between the member countries. Becoming a member of an existing union usually requires the approval of both the candidate and members, while leaving it may require only the unilateral will of the exiting country. There are many examples of accession of states to previously consolidated economic unions, and a recent example of leaving is the withdrawal of the United Kingdom from the European Union. Motivated by the Brexit process, in this paper we propose an agent-based model to study the determinant factors driving withdrawals from an economic union. We show that both Union and local taxes promote the exits, whereas customs fees out of the Union boost cohesion. Furthermore, heterogeneity in both business conditions and wealth distribution promotes withdrawals, while countries' size diversity does not have a significant effect on them. We also deep into the individual causes that lead to dissatisfaction and, ultimately, to exits. We found that, for low Union taxes, the wealth inequality within the country is the leading cause of anti-Union opinion spreading. Conversely, for high Union taxes, the country's performance turns out to be the main driving force, resulting in a risk of wealthier countries leaving the Union. These findings will be helpful for the design of economic policies and effective informative campaigns.

\newpage
\section{Introduction}

Economic unions are international blocs with a high degree of economic integration, constituting a common market through customs removal and a shared external trade policy \cite{grossman2016purpose}. Furthermore, they show a high degree of freedom in the movement of goods and services, and participant states share policies on product regulation and production factors. Examples of economic unions (to a greater or lesser extent of implementation) are the European Union \cite{kiljunen2004european}, the CARICOM Single Market and Economy \cite{berry2014caribbean}, the Central American Common Market \cite{schooler1965product,bulmer1998central}, the Eurasian Economic Union, \cite{vinokurov2017eurasian} and the Gulf Cooperation Council \cite{alqahtani2018financial}.

Once an economic union is constituted, there may be both adhesion and withdrawal processes, being the European Union enlargement \cite{schimmelfennig2005politics,borzel2017european,alesina2017europe,campos2019institutional,campos2022institutional} and the withdrawal of the United Kingdom (Brexit) \cite{clarke2017brexit,arnorsson2018causes}, respectively, recent examples. In this regard, as being part of a union requires a certain degree of acceptance by the citizens, especially for democracies, it is crucial to know how different factors affect the opinion about the Union, both among Union citizens and in accession candidate states \cite{dijkstra2020geography}. Mathematical modeling constitutes a powerful approach to study both economic \cite{hritonenko1999mathematical,naldi2010mathematical} and social systems \cite{castellano2009statistical}. On the one hand, economic and trade aspects of economic unions have been studied through theoretical models \cite{alesina2005international,lorz2013size,gancia2020theory,alesina2000economic}. On the other hand, social dynamics \cite{tuma1984social} and, in particular, opinion diffusion \cite{sirbu2017opinion} constitutes a field of study in which agent-based models have demonstrated their efficacy in qualitatively explaining collective behavior \cite{clifford1973model,axelrod1997dissemination,fernandez2014voter,gracia2009residential,hernandez2018robustness}. 

In this paper, we propose a socio-economic agent-based model that incorporates both opinion exchange and economic ingredients to study the feedback between the trading aspects of an economic union and the individual opinions about it. The main goal of this work is to characterize a set of plausible factors that affect the desire of citizens to be part of the Union and, by extension, the dynamics of accessions and withdrawals from it. The model here presented replicates an economic union among societies (hereafter countries, although it may refer to countries, sovereign states, or entities with political and economic independence and decision power). To this end, the dynamic includes three levels: i) a macro level that accounts for the interactions among countries in and out of the Union, such as Union taxes, customs fees, accessions and withdrawals to the Union, as well as Union taxes redistribution among member countries ii) a mesoscale level that regulates the interactions between agents and their country, including local taxes and their redistribution among the citizens, and iii) a micro-level that accounts for the interactions among the agents (i.e., the citizens). The interactions among agents will be of both trading and opinion exchange kinds. Ultimately, feedback between the macro-, meso-, and the micro-level will determine the dynamic of the system and the robustness of the Union. 

By performing extensive numerical simulations, we identify the main factors that affect the dynamics of the Union, including the evolution of the agents' opinion about the Union and the countries accessions to and withdrawals from it. We show that an increment in either Union or local taxes results in a higher risk of withdrawals, being the integrity very sensitive to Union taxes rises. Regarding the individual-level causes for the spreading of dissatisfaction with the Union and the subsequent possible withdrawals, we show that taxes play a key role in the anti-Union opinion spreading: For low Union and local taxes, the differences in wealth within countries result to be the main cause of Union exits, while as Union taxes increase, differences among countries gains influence. Furthermore, we checked these results under different scenarios such as unequal initial wealth distribution, heterogeneous business conditions among countries, and dissimilar country sizes. We find that heterogeneity in both initial wealth distribution and business conditions increases the risk of withdrawals. Contrarily, there is not a significant effect of countries' size distribution on withdrawals' risk.

\section{The model}
\label{theModel}

\subsection*{Agents, countries}
\begin{figure}[ht!]
 \includegraphics[width=\columnwidth]{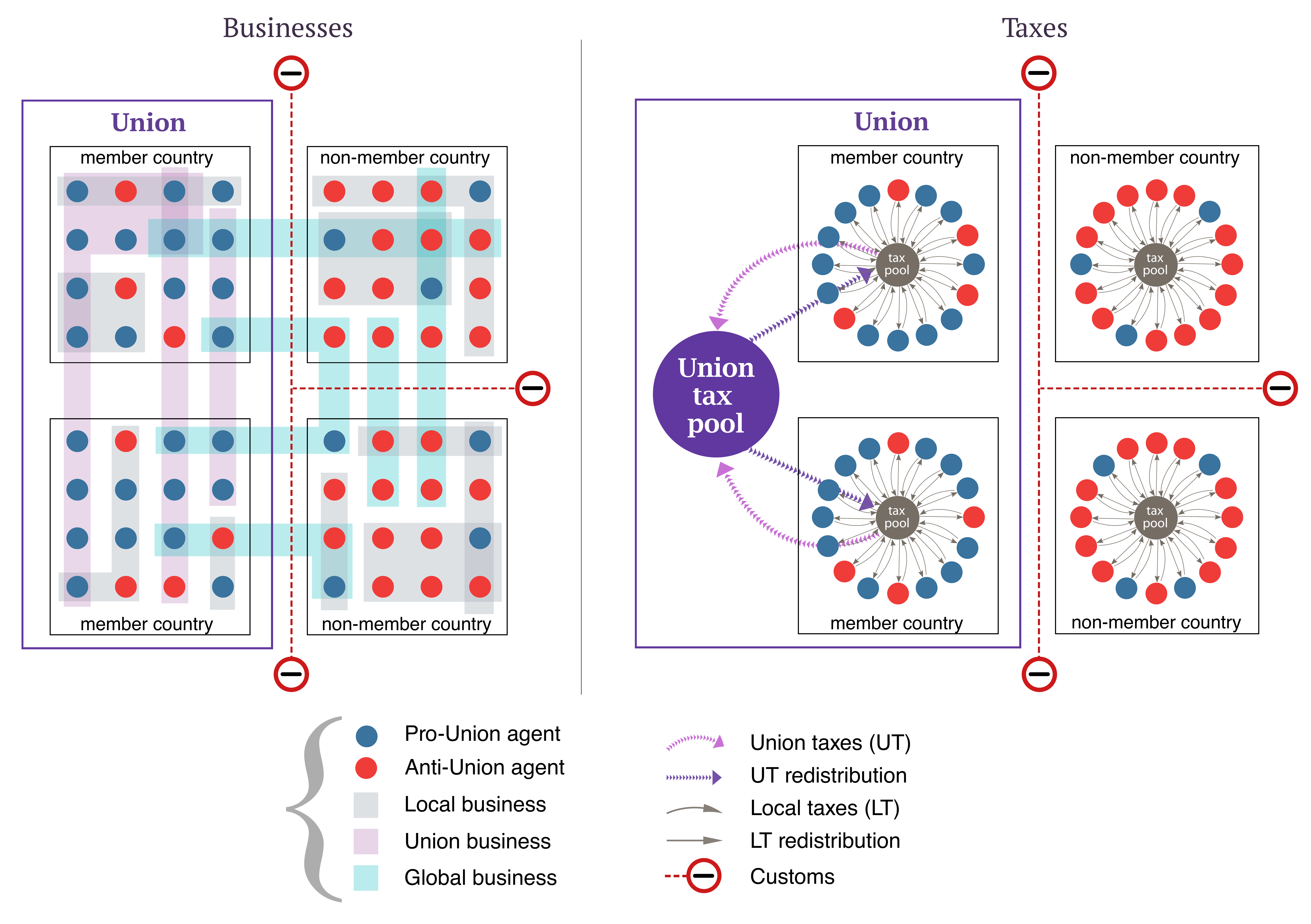} 
 \caption{\textbf{Schematic representation of the model.} 
 Left panel (\textbf{Businesses}): Agents (which represent citizens) are located in countries. Some countries belong to an economic Union. Agents are involved in different businesses. There are three kinds of businesses: i) local, restricted to a country, ii) Union, restricted to Union agents, and iii) global businesses. Right panel (\textbf{Taxes}): Agents pay local taxes according to their gross incomes. Those taxes are equally redistributed, together with other countries' incomes as custom taxes, after being multiplied by an enhancement factor. Furthermore, Union members contribute to the Union proportionally to their gross incomes. Union incomes, after enhancement, are redistributed among the members inversely proportional to countries' wealth. Agents may change their opinion about the Union either by imitation or by social pressure. Withdrawals (resp., adhesions) may occur when the anti-Union (pro-Union) rate into a country exceeds a threshold for a long enough time.}
\label{fig1}
\end{figure}

Fig. \ref{fig1} displays a schematic representation of the model, and Table \ref{tabparameters} a parameters summary. We consider an economic Union and a set of $C$ countries. Each country $i$ ($i=1,2,\ldots,C$) is modeled by:
\begin{itemize}
    \item A population of $N_i$ agents, i.e., its citizens.
    \item A flag $U_i$ that says whether the country is part of the union ($U_i=1$) or not ($U_i=0$).
\end{itemize}

Each agent $a_i^j$ ($i=1, 2,\ldots, C; j=1, 2, \ldots, N_i$) has associated a value $\sigma_i^j\in\{0,1\}$ that determines her opinion about the membership of her country to the Union. Specifically, each agent $a_i^j$ is either pro-Union ($\sigma_i^j=1$) or anti-Union ($\sigma_i^j=0$).

\subsection*{Business}

Agents are repeatedly involved in private enterprises/businesses modeled as games. There are a total of $B$ businesses, distributed into three kinds:

\begin{itemize}
    \item \textbf{Local businesses (LB)}, which are restricted to agents within a country $i$.
    
    The LB \textit{per capita} profit is given by: 
    \begin{equation}
        Y_{L,i}=\Gamma_{L,i}f(n)-1\;\; .
        \label{LBProfit}
    \end{equation}
    Here, it is assumed an individual contribution per round of 1. $\Gamma_{L,i}$ refers to the $i$-country’s local business’ profitability, and $f(n)$ to an activation function in $[0,1]$ that takes into account scale economies so that the full profitability of the business is obtained only if the number of partners exceeds a certain critical mass $x$.
    \begin{equation}
        f(n)=\frac{1}{1+(x/n)^\beta}\;\; , 
        \label{f_function}
    \end{equation}
    where $n>0$ is the number of agents participating in the business, $x$ refers to the critical mass (for $n=x$, the yield is half the maximum), and $\beta$ is the Hill coefficient: as $\beta$ is increased, the yield curve becomes steeper.
    The individual payoff from a LB is given by:
    \begin{equation}
    \Pi_{LB,i} =
     \begin{cases}
      Y_{L,i}(1-T_i) & \text{if }\; Y_{L,i}>0 \;\; ,\\
      Y_{L,i} & \text{otherwise},
     \end{cases}
    \label{LBPayoff}
    \end{equation}
    where $T_i$ refers to the $i$-country’s average tax rate (local taxes). 
    
    \item \textbf{Union businesses (UB)}, restricted to agents within the union.
    
    Similarly to with Eq. \ref{LBProfit}, the \textit{per capita} profit of a UB is given by:
    \begin{equation}
        Y_{U}=\Gamma_{U}f(n)-1,
        \label{UBProfit}
    \end{equation}
    where $\Gamma_U$ refers to the Union business’ profitability. Like in the case of LB (Eq. \ref{LBPayoff}), the individual payoff from a UB is given by:
    
    \begin{equation}
    \Pi_{UB,i} =
    \begin{cases}
      Y_{U}(1-T_i) & \text{if }\; Y_{U}>0 \;\; ,\\
      Y_{U} & \text{otherwise}.
    \end{cases}
    \label{UBPayoff}
    \end{equation}
    
    Note that the only differences between LBs and UBs are the profitability (in the cases in which it differs) and the fact that LBs are restricted to agents of a given country, whereas UBs are open to any agent within the Union.

    \item \textbf{Global businesses (GB)}, which are open to all the agents. They involve customs fees. 

    As in previous cases, the GB \textit{per capita} profit is given by:
    \begin{equation}
        Y_{G}=\Gamma_{G}f(n)-1\;\; ,
        \label{GBProfit}
    \end{equation}
    where $\Gamma_{G}$ refers to the global business’ profitability. The GB individual Payoff is:
    
    \begin{equation}
     \Pi_{GB,i} =
     \begin{cases}
      Y_{G}(1-T_i-C_iS_{i,u}) & \text{if }\; Y_{G}>0 \;\; ,\\
      Y_{G} & \text{otherwise},
     \end{cases}
     \label{GBPayoff}
    \end{equation}
    where $C_i$ refers to the customs fees, $u$ ($u=1,2,\ldots$) stands for the global business, and $S_{i,u}$ represents the fraction of i) no-Union partners in the GB $u$ if agent $i$ belongs to the Union, or ii) foreigner partners in $u$ if $i$ does not belong to the Union. The term $C_iS_i$ models the fact that incomes, goods, and services from no-Union countries pay customs fees, 
    being the Union perimeter a border for this purpose.
\end{itemize}


\subsection*{Taxes}

Local taxes are taken from the businesses by the country $i$ as per its own rate $T_i$ (from Eqs. \ref{LBPayoff}, \ref{UBPayoff}, and \ref{GBPayoff}, agents pay local taxes according to their gross incomes and their country's rate). Furthermore, if the country belongs to the Union, it pays a Union tax proportional to the instant gross profits of its agents (i.e., the positive profits $Y_{L,i},Y_{U},Y_{G}$ obtained in the last round); that amount is discounted from the income that the country has received from taxes and customs. Union taxes are redistributed among all the countries of the Union inversely proportional to the accumulated payoff over time of all the agents in each country, i.e., poor countries receive more Union’s incomes than rich ones. Finally, country incomes (taxes + custom - Union taxes + Union redistribution) are multiplied by an enhancement factor and equally shared among all the agents of the country. All these operations and the associated parameters will be described in detail in the Dynamics subsection.

Note that in the model, the taxes and their redistribution system represent duties, public incomes, infrastructures and public goods, as well as the costs associated with institutions (including the Union) and their maintenance.

\subsection*{Agents' actions}

The agents change their opinion from pro- to anti-Union and vice versa according to two mechanisms. The first mechanism is based on opinion spreading and is driven by both the social pressure and the dissatisfaction or satisfaction produced by the balance of payments between the country and the Union. The second mechanism relies on the imitation of the agents with better performance. Countries enter and leave the Union when their fractions of pro- and anti-Union agents exceed respective thresholds for a sufficiently long time. The Dynamics subsection contains the details of both mechanisms.

\subsection*{Initial Conditions}

At the beginning of each simulation, each country $i$ has a probability $P(U_i=1)$ of belonging to the Union. Unless otherwise stated, $P(U_i=1)=1/2\;, \forall i$.

Businesses are designed as local or non-local with a probability $\xi_L$. Explicitly, each business has a probability $\xi_L$ of being LB, a probability $\xi_U$ of being UB and, therefore, a probability $\xi_G=1-\xi_L-\xi_U$ of being GB. Unless otherwise stated,  $\xi_L=1/2$, $\xi_U=1/2\times1/3=1/6$ (then, $\xi_G=1/2\times2/3=2/6$). LBs are
distributed among all the countries $i=1,2,\ldots,C$ with probability proportional to their population $N_i$.

Each agent belonging (resp., non belonging) to the Union has a probability $p_0$ (resp., $p'_0$) of being pro-Union. Unless otherwise stated,  $p_0=0.6$, $p'_0=0.4$.

Each agent is assigned to different businesses in accordance with the preferences dictated by her pro- or anti-Union opinion, that is, for agents belonging to the Union, pro-Union ones participate more likely in Union businesses than in global ones. Nevertheless, the expected number of businesses in which each agent participates is equal for all the agents, i.e., it does not depend on her belonging to the Union nor on her opinion. Explicitly, all businesses are assumed to have on average $x$ partners, where $x$ refers to the critical participation threshold in Eq. \ref{f_function}.

\begin{itemize}
    \item LB memberships are distributed among all the agents of the country to which the business is associated. Each agent $a_i^j$ of a given country $i$ has the same probability $x/N_i$ to participate in each LB located in country $i$, so that each LB has on average $x$ partners. 
    \item UB memberships are distributed among all the agents of the Union. Each pro-Union agent $a_i^j$ of the Union ($\sigma_i^j=1,U_i=1$) has a probability $2\nu$ to participate in each UB, being that probability half ($\nu$) for anti-Union agents.
    The condition that the expected number of partners per UB is equal to $x$ can be written as
\begin{equation}
    2 N_U p_0 \nu + N_U(1-p_0)\nu=x\,,\quad N_U=\sum_{i=1}^CN_i U_i\,,\\
    \label{expectedUB}
\end{equation}
where $N_U$ is the initial number of agents in the union, $N_Up_0$ and $N_U(1-p_0)$ of which are pro- and anti-Union, respectively. From Eq. \ref{expectedUB}, it follows:
\begin{eqnarray}
    \label{UB_IC_formula}
    \nu = \frac{x}{N_U(p_0+1)}\,.
\end{eqnarray}

    \item GB memberships are distributed among all the agents of the system. Let $\omega_o$ be the initial probability of participating in a GB for agents out of the Union, $\omega_p$ for pro-Union agents in the Union, and $\omega_a$ for anti-Union agents in the Union.
    First, we equalize the expected number of business participation for all the agents, in and out of the Union, i.e., we impose:
\begin{equation}
    \xi_G\omega_o=\xi_G\omega_p+2\xi_U\nu=\xi_G\omega_a+\xi_U\nu\;,
    \label{equalizingGB}
\end{equation}
where $2\nu$ and $\nu$ are, respectively, the probabilities for a pro- and anti-Union agent in the Union to participate in a given UB (see Eq. \ref{UB_IC_formula}).
Then, similarly to Eq. \ref{expectedUB}, the condition on the expected number of partners per GB can be written as
\begin{equation}
    N_o\omega_o+N_Up_0\omega_p+N_U(1-p_0)\omega_a=x\,,\quad
    N_o=\sum_{i=1}^CN_i (1-U_i)\,,\\
    \label{expectedGB}
\end{equation}
where $N_o$ is the initial number of agents out of the union.

 For plausible conditions (such as $\xi_G,N_U>0 ;\;\xi_G\omega_0>2\xi_U\nu$), the positive solutions of the system of Eqs. \ref{equalizingGB}\,-\ref{expectedGB} provide the GB participation probabilities:
 \begin{eqnarray}
    \omega_o&=& \frac{x+N_U\phi\nu(p_0+1)}{N}  \nonumber\;,\\
    \omega_p&=& \omega_o - 2 \nu \phi \nonumber\;,\\
    \omega_a&=& \omega_o - \nu \phi  \nonumber\;,\\
    \label{GB_IC_formula}
    \label{GB_IC_formula}
\end{eqnarray}
where $\phi =\xi_U/\xi_G$ and $N$ is the total number of agents: $N=N_o+N_U$. 
\end{itemize}


Concerning initial payoffs, we consider two scenarios:

\begin{itemize}
    \item In the homogeneous scenario, all agents in the system are initially provided with the same accumulated payoff: $W_i^j(t=0)=10$, $\forall i,j$.
    \item In the heterogeneous scenario, some countries are richer than others. Furthermore, within each country, some agents are richer than others. For each country $i$, agents are initially provided with an accumulated payoff $W_i^j(t=0)$ according to a Poisson distribution $P(\lambda_i)$. The $\lambda_i$ values (which represent countries’ wealth) follow a normal distribution ($\mu=10$, $\sigma^2=4$)\footnote{Given the model constraints, a Poisson-normal distribution is chosen to allow the initial assignation of businesses. Nevertheless, the dynamic will lead the system to a heavy-tailed (Pareto-like) wealth distribution.}.
\end{itemize}

\subsection*{Dynamics}
Let us define the dynamics of the model. Each synchronous time step (i.e., each round) corresponds to a week, a period in which agents interact through businesses and exchange opinions among them. At each round:
\begin{enumerate}
    \item Each agent $a_i^j$ participates in all the businesses she is involved in and obtain the correspondent profits according to Eqs. \ref{LBProfit}, \ref{UBProfit}, and \ref{GBProfit}.
    \begin{itemize}
        \item For each business with a positive profit each agent $a_i^j$ participates in, a fraction $T_i$ of the profit is discounted and collected in the agent’s country common pool $\Phi_i$ (Eqs. \ref{LBPayoff}, \ref{UBPayoff}, and \ref{GBPayoff}).
        \item Furthermore, for each GB business she participates in, a fraction $C_i S_{i,u}$ 
        of the profit (if positive) is discounted as customs and collected in the agent’s country common pool $\Phi_i$ (Eq. \ref{UBPayoff}). 
    \end{itemize}

    \item The instant payoff $\Pi_i^j$ of each agent $a_i^j$ is the remaining from the profits after local taxes and customs are discounted according to Eqs. \ref{LBPayoff}, \ref{UBPayoff}, and  \ref{GBPayoff}.
    \item For countries belonging to the Union ($U_i=1$), a fraction $T_u$ of all individual positive profits is discounted from the corresponding countries common pools $\Phi_i$ and collected in the Union’s common pool $\Phi(\mathrm{U})$.
    \item The Union's common pool amount $\Phi(\mathrm{U})$ is redistributed among all the countries common pools $\Phi_i$ ($\forall i \mid U_i=1$) of the Union inversely proportional to the accumulated payoff over time of all the agents in each country, i.e., poor countries receive more Union incomes than rich ones.
    \item For each country $i=1,2,\ldots,C$, its common pool $\Phi_i$ is multiplied by an enhancement factor $\alpha$ and, subsequently, equally distributed among all the agents $a_i^j, j=1,2;\ldots,N_i$.
    \item Agents’ accumulated payoffs $W_i^j$ (agents' wealth) evolve over time according to:
    \begin{equation}
    W_i^j(t+1)= \Pi_i^j(t)+ \max(\rho W_i^j(t)-1, 0),
    \end{equation}
    where $\Pi_i ^ j (t) $ is the instant payoff obtained in the current round and $\rho $ is a parameter that accounts for both inflation-deflation and scalable expenses, while the subtracted unit sets the minimum life-expense.
    \item Opinion dynamics.
    Agents change their opinions about the Union based on social pressure driven by i) the country-Union economic balance (that is, how advantageous membership is in the short term), and ii) the agents' influence: the wealthier an agent is, the more influential. 
    \begin{itemize}
        \item For each pro-Union agent $a_i^j$ belonging to the Union ($\sigma_i^j=U_i=1$), there is a probability $P_i(p\to a)$ to become anti-Union ($\sigma_i^j:\;1\to 0$), according to a function of the difference between the contributions by its country $i$ to the Union and the benefits that are obtained from the Union:

        \begin{eqnarray} \label{OpinionDynamics1}
          P_i(p \to a)&=&
          \begin{cases}
           \epsilon \Delta_i (1+ g_i) & \text{if }\; \Delta_i >0 \;\; ,\\
           0 & \text{otherwise},
          \end{cases}
        \end{eqnarray}
        with:\\
        \begin{eqnarray}
          \Delta_i&=& \frac{C_i-R_i}{C_i},\nonumber\\
          g_i&=&\frac{\sum_j[(1-\sigma_i^j)(W_i^j)^\eta]}{\sum_j(W_i^j)^\eta},\nonumber
        \end{eqnarray}
        where $C_i$ is the $i$-country contribution to the Union, $R_i$ is the received subside (i.e., the amount that country $i$ receives from the Union through tax redistribution by the Union) and $\Delta_i$ is the unbalance normalized over the $i$-country contribution. $g_i$ is the weighted fraction of anti-Union agents in the country $i$, where each agent has been weighted by his accumulated payoff $W_i^j$ raised to $\eta$. Here, the exponent $\eta \geq 0$ tunes the effect of wealth on propaganda, for $\eta=0$ all the agents have the same effect regardless of their accumulated payoff; the higher $\eta$, the more relatively influential are rich agents. $\epsilon$ modulates the specific influence of the opinion dynamics on the system dynamics: $0<\epsilon<0.5$.
        
        \item Correspondingly, for each anti-Union agent belonging to the Union ($a_i^j:\; \sigma_i^j=0, U_i=1$) there is a probability $P_i(a \to p)$ to become pro-Union:
        
        \begin{eqnarray} \label{OpinionDynamics2}
          P_i(a \to p)&=&
          \begin{cases}
           -\epsilon \Delta_i (1+ g'_i) & \text{if }\; \Delta_i <0 \;\; ,\\
           0 & \text{otherwise},
          \end{cases}
        \end{eqnarray}
        with:\\
        \begin{eqnarray}
          g'_i&=&\frac{\sum_j[\sigma_i^j(W_i^j)^\eta]}{\sum_ j(W_i^j)^\eta}\;\; \nonumber
        \end{eqnarray}
        

        \item Previous probabilities given by Eqs. (\ref{OpinionDynamics1}-\ref{OpinionDynamics2}) also apply to agents not belonging to the Union. In that case, $C_i$ , $R_i$, $\Delta_i$ and $g_i$ are computed over the country $i$ that is closest to the non-union agent in terms of \textit{per capita} wealth (averaged accumulated payoff).
        \end{itemize}
        \item Imitation dynamics:
        \begin{itemize}
            \item Agents that have not updated their strategy in the current step are allowed to change strategy (from pro- to anti-Union and vice versa) by imitation. Each agent $a_i^j$ will randomly choose an agent $a_z^k$ from his country ($i=z$) with probability $\psi$, otherwise (i.e., with a probability $1-\psi$) from the whole system. Then, agent $a_i^j$ will adopt the agent $a_z^k$ strategy with a probability given by the normalized payoffs' difference between both agents, if positive:
 
           \begin{eqnarray}
             P(\sigma_i^j \leftarrow \sigma_z^k) &=& 
             \begin{cases}
              \frac{W_z^k - W_i^j}{W_z^k} & \text{if }\; W_z^k> W_i^j,\\
              0 & \text{otherwise}.
             \end{cases}
            \end{eqnarray}
        \end{itemize}
        \item Each agent tries to be part of a business she is not involved in. The tentative business to be part of is chosen proportionally to business' profits. The probability of been accepted in a business is proportional to the accumulated payoff of the applicant. Applicants with null or negative accumulated payoffs are not accepted. If the agent is accepted, she leaves the business with the lower payoff among those she is involved in, provided it is lower than the expected payoff from the business she applied for (otherwise, nothing happens).
        \item A country of the Union remains in the Union as long as its fraction of pro-Union agents is above 45\%. If at a given step, the fraction of pro-Union agents is below 45\% and, subsequently, it remains below 50\% for 20 consecutive steps, the country leaves the Union. Agents of countries leaving the Union will quit Union businesses. 
        For each Union business an agent leaves, in the case of having positive accumulated payoff, she tries to be part of a global business she is not involved in. The tentative business to be part of is chosen proportionally to business' profits. The probability of being accepted in the business is proportional to the individual accumulated payoff of the applicant. If not accepted, she will try again (by choosing the new business proportionally to its yield) until accepted.
        \item A country not belonging to the Union incorporates to it when the fraction of pro-Union agents at a given step is above 55\% and, subsequently, it remains above 50\% for 20 consecutive steps. Agents entering the Union will be allowed to participate in Union Businesses in subsequent steps.

\end{enumerate}

\begin{table}[]
\centering
 \begin{tabular}{ |p{1.2cm}|p{7.8cm}|p{2.7cm}|  }
  \hline
  \multicolumn{3}{|c|}{Parameters} \\
  \hline
  symbol & description & default (range)\\
  \hline
  $C$ & number of countries & 20 \\
  $N_i$ & agents per country & $10^5$ ($10^4$\;-\;4x$10^5$)\\
  $\Gamma_L$ & local businesses' profitability & 1.5 (1.3\;-\;1.7)\\
  $\Gamma_U$ & Union businesses' profitability & 1.5 \\
  $\Gamma_G$ & global businesses' profitability & 1.5 \\
  $\beta$ & hardness of the business activation function &  1.0\\
  $\alpha$ & common pool's enhancement factor & 1.01 \\
  $T_i$ & local taxes & 0.2 (0.1\;-\;0.5)\\
  $T_U$ & Union taxes & 0.05 (0\;-\;0.1)\\
  $C_i$ & customs fees & 0.1 (0.1\;-\;0.2)\\
  $\eta$ & effect of wealth on propaganda & 1.0 \\
  $\epsilon$ & influence of opinion dynamics & 0.1 \\
  $\rho$ & inflation-deflation and expenses besides minimum &  0.95\\
  \hline
  $\xi_L$ & initial ratio of local businesses & 1/2 \\
  $\xi_U$ & initial ratio of Union businesses & 1/6 \\
  $p_0$ & initial ratio of pro-Union agents in the Union & $3/5$ \\
  $p'_0$ & initial ratio of pro-Union agents out of the Union & $2/5$ \\
  \hline
 \end{tabular}
 \caption{Model's parameters with their symbol, description, and default value or range taken in the simulations. Given the arbitrariness of some choices, robustness analyses were performed by varying the parameters' values: In some simulations, the values differ from the default ones, as indicated in the text and captions. Note that a businesses' profitability of 1.5, although representing 50\% per week, yields, on average, a yearly \textit{per-capita} profit of $\sim50\%$.} 
 \label{tabparameters}
\end{table}

\newpage

\section{Results and discussion}

\begin{figure}
 \includegraphics[width=\columnwidth]{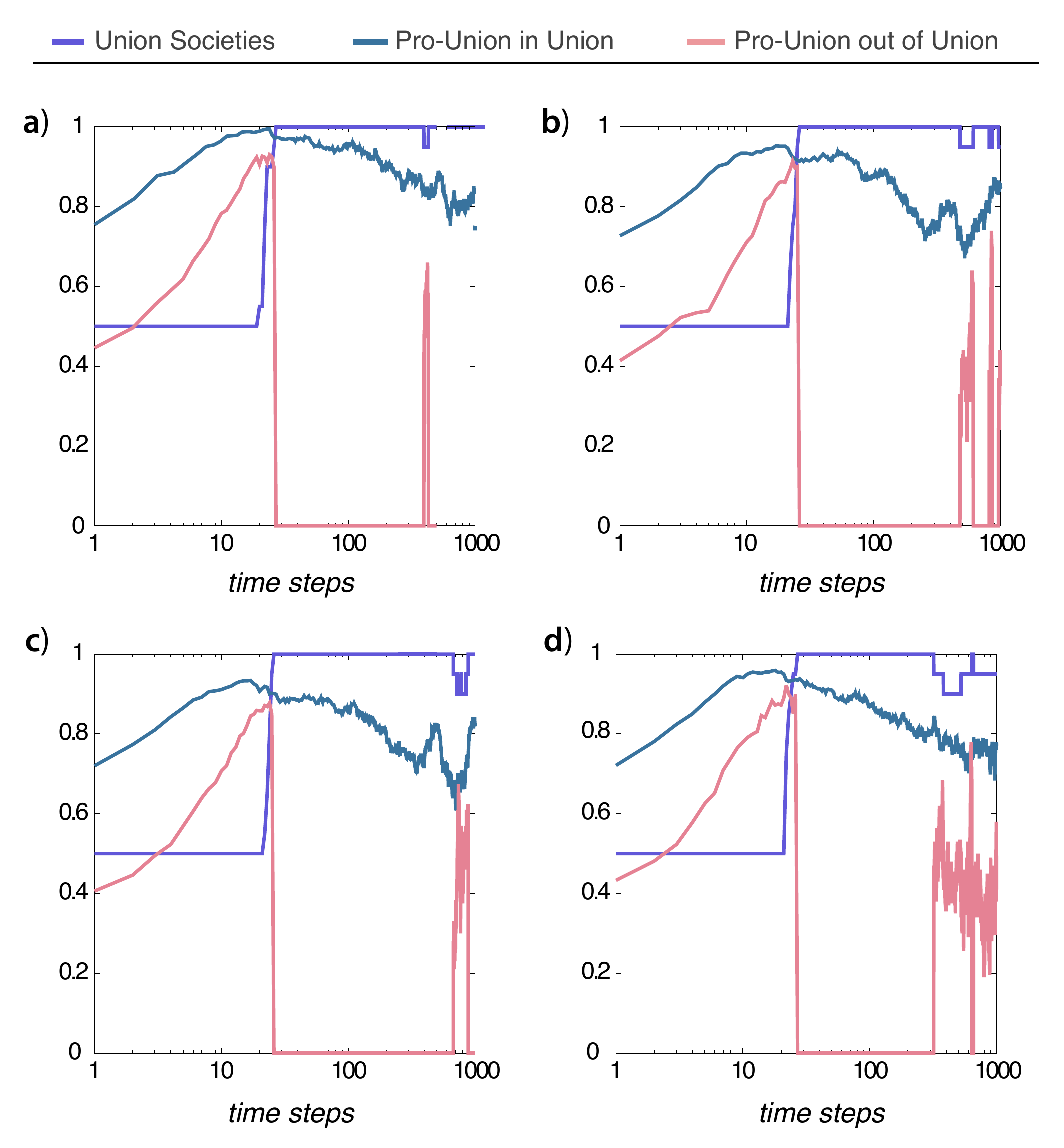} 
 \caption{\textbf{Characteristic dynamics.} Time evolution for the fraction of Union countries (purple), Pro-union agents in the Union (blue), and Pro-union agents out of the Union (orange). Each panel corresponds to a characteristic realization for a value of the Union tax rate: $T_U=0.1$, $0.05$, $0.025$, and $0.01$ in panels \textbf{A)}, \textbf{B)}, \textbf{C)}, and \textbf{D)}, respectively. The other parameters are: $C=20$, $N_i=10^5$, $T_i=0.3$, $C_i=0.1$, $\Gamma_L=\Gamma_U=\Gamma_G=1.5$, $\psi=0.8$, $\rho=0.95$, $x=20$, $B=5\times10^{-4}$, $\epsilon=0.1$, and $\eta=1$.
 }
\label{fig2}
\end{figure}

We begin the analysis of our results with a description of the temporal evolution of both the accessions and withdrawals from the Union and the opinion that citizens have about belonging to it. Figure \ref{fig2} shows the fraction of countries within the Union (purple line) along with the ratio of pro-Union agents in (blue) and out (orange) of the Union, as a function of time. Each panel corresponds to a characteristic realization 
for decreasing values of the Union tax rate, A) to D).
As shown, there is an initial transition period in which pro-Union opinion becomes dominant and all the countries join the Union. After this transient, opinion about the membership to the Union fluctuates, and occasional Union withdrawals (exits) may occur.

\subsection*{Withdrawals}

\begin{figure}
 \includegraphics[width=\columnwidth]{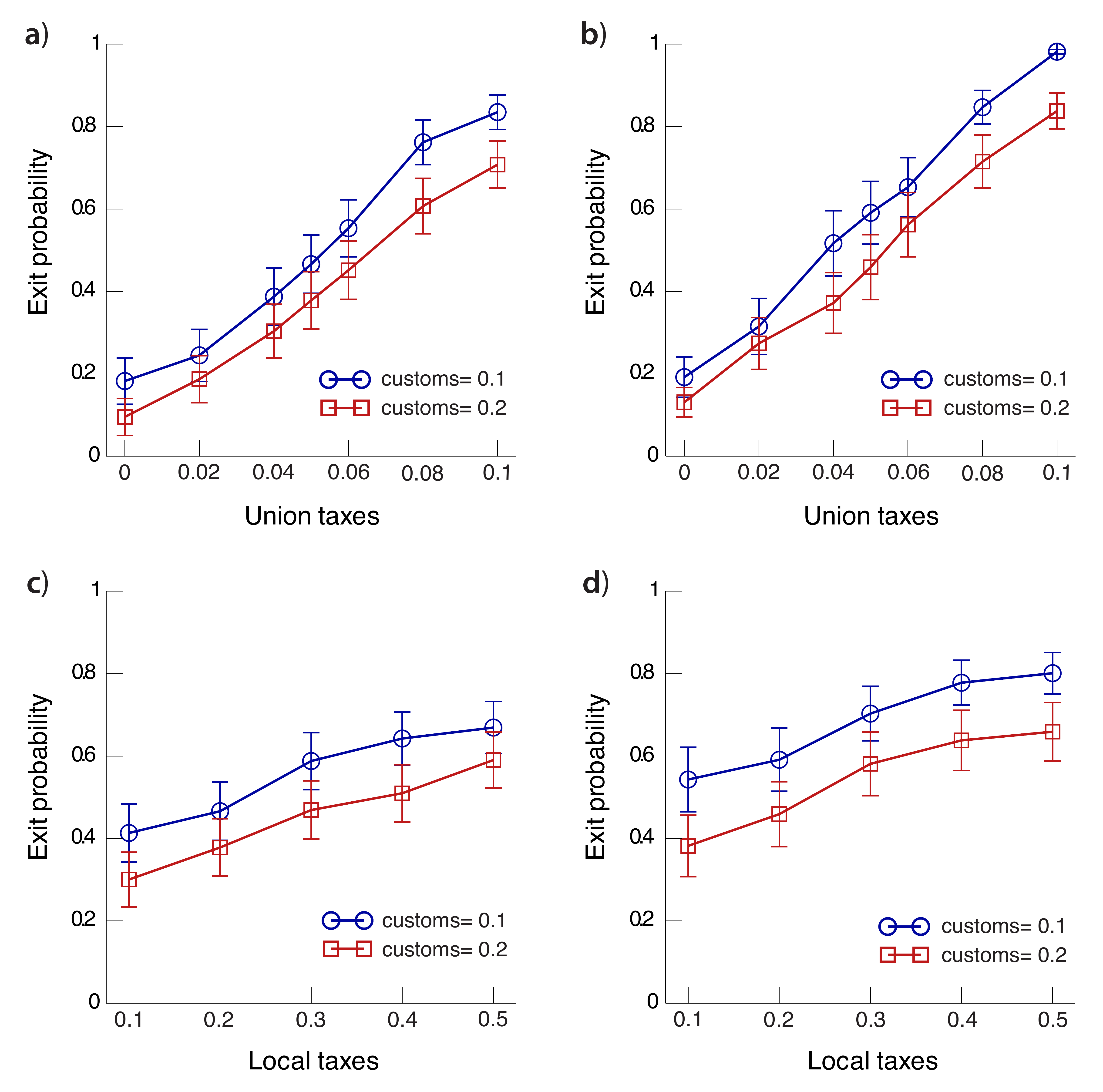} 
 \caption{\textbf{Withdrawal probability in the long-term versus taxes.} Probability of at least an exit event (i.e., a country leaves the Union) in a finite interval of 1,000 rounds ($\sim20$ years). Top panels \textbf{A}-\textbf{B} show the results as a function of the Union tax rate $T_U$ for a fixed local tax rate of $T_i=0.2$. Bottom panels \textbf{C}-\textbf{D} show the results versus the local tax rate $T_i$ for a fixed Union tax rate of $T_U=0.05$. Left panels \textbf{A,C} display the homogeneous scenario: All agents in the system have the same accumulated payoff at $t=0$. Right panels \textbf{B,D)} display the heterogeneous scenario for the initial conditions: For each country $i$, agents' initial wealth follows a Poisson distribution $P(\lambda_i)$, the $P(\lambda_i)$ values are taking from a normal distribution ($\mu=10$, $\sigma^2=4$). In all the panels, $C=20$, $N_i=10^5$, $\Gamma_L=\Gamma_U=\Gamma_G=1.5$, $\psi=0.8$, $\rho=0.95$, $x=20$, $B=5\times10^{-4}$, $\epsilon=0.1$, and $\eta=1$. Error bars represent standard error of the mean (SEM). Each point is averaged over 10,000  numerical simulations.}
 \label{fig3}
\end{figure}

To study in-depth the influence of taxation policy on withdrawals from the Union, we focus on the probability of an exit event (i.e., that a country leaves the Union) in 1,000 rounds (which, in the model, corresponds to $\sim20$ years) as a function of Union taxes in different scenarios. Specifically, within each simulation, we analyzed a time window of 1000 steps after a transient of 1000 steps. Each point corresponds to 1000 independent random sets of initial conditions and 10 realizations for each set of initial conditions, constituting 10,000 realizations per point. Panels A-B of Fig. \ref{fig3} show the probability of having, at least, a withdrawal from the Union along 1000 steps versus the Union tax rate $T_U$ for a fixed local tax rate of $T_i=0.2$. Blue circles correspond to low customs fees ($C_i=0.1$) and red squares to high customs ($C_i=0.2$). Panel A of Fig. \ref{fig3} displays the results for the homogeneous scenario in which all the agents start with the same initial endorsement, mimicking the far-from-realistic situation of a homogeneous distribution of \textit{per capita} wealth among citizens and countries. As shown, the withdrawal probability increases with Union taxes and decreases with customs. To check the role of wealth heterogeneity on exits, Panel B of Fig. \ref{fig3} shows the same probability for the heterogeneous scenario where initial endorsements are unequally distributed among countries and agents (see previous Initial Conditions section). This more realistic wealth distribution shows the same qualitative dependency of exits with Union taxes and customs fees as the homogeneous one. Furthermore, a quantitative comparison of panels A and B highlights that heterogeneity in the initial conditions increases the exit probability: When all the countries present the same initial wealth (Panel A), \textit{ceteris paribus}, the exit probability is lower than in the heterogeneous scenario, where i) some countries have a higher \textit{per capita} wealth than others, and ii) wealth is unequally distributed within the countries (Panel B). 

\begin{figure}
 \includegraphics[width=\columnwidth]{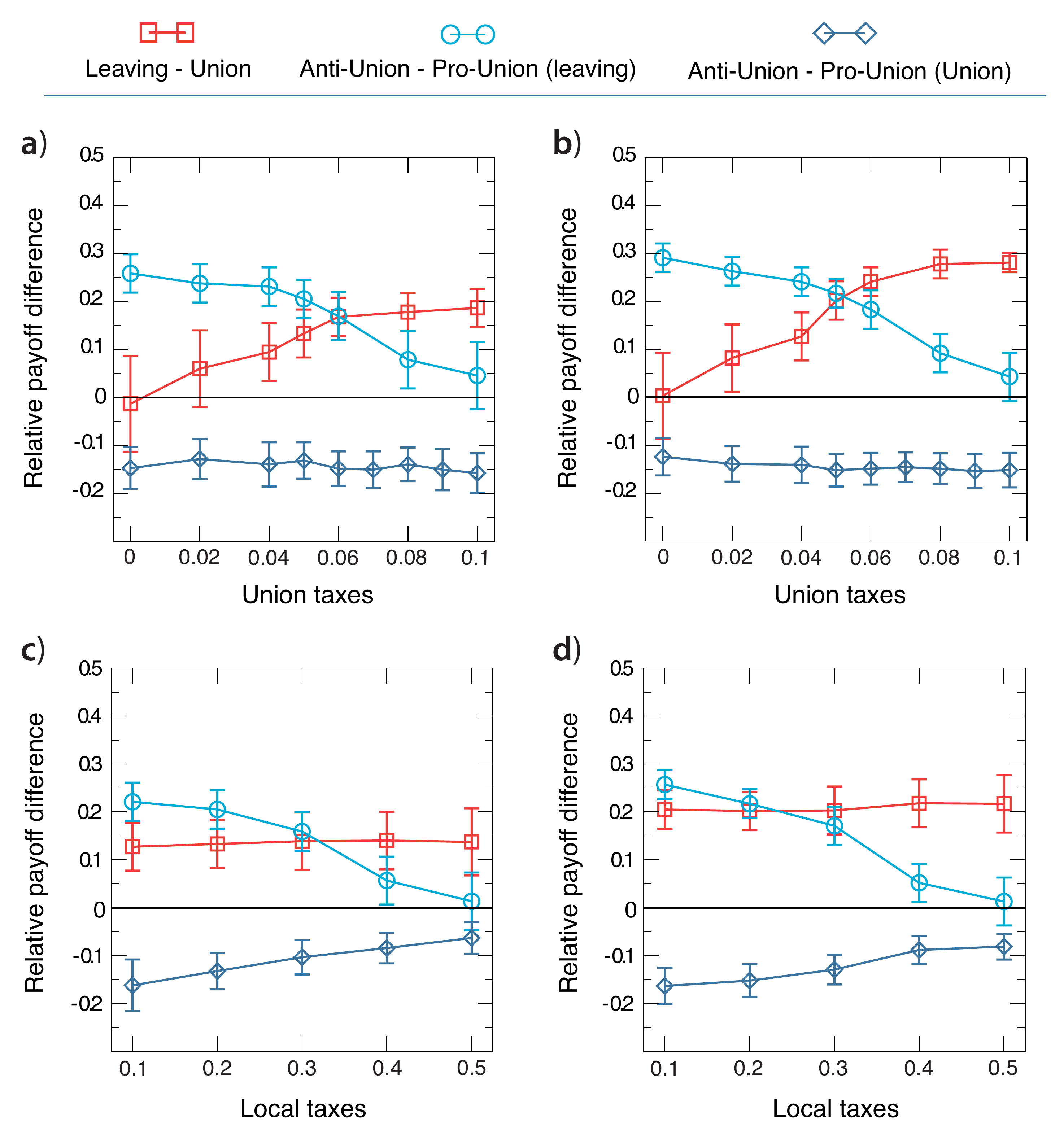} 
 \caption{\textbf{Withdrawals' drivers for different local and Union taxation}. Payoff differences at the moment of an exit as a function of the Union tax rate $T_U$ (top panels) and versus the local tax rate $T_i$ (bottom). Red squares represent the relative difference between the accumulated payoff in the country that leaves the Union and the average accumulated payoff in the Union at the moment of an exit. Blue circles (resp., diamonds) represent, at the moment of an exit, the difference between the average accumulated payoff of anti-Union agents and pro-Union agents in the country leaving the Union (resp., in the Union). Left panels \textbf{A,C} correspond to the homogeneous scenario: All agents in the system have the same accumulated payoff at $t=0$. Right panels \textbf{B,D)} correspond to the heterogeneous scenario for the initial conditions: For each country $i$, agents' initial wealth follows a Poisson distribution $P(\lambda_i)$, the $\lambda_i$ values are taking from a normal distribution ($\mu=10$, $\sigma^2=4$). In all the panels, $C=20$, $N_i=10^5$, $\Gamma_L=\Gamma_U=\Gamma_G=1.5$, $\psi=0.8$, $\rho=0.95$, $x=20$, $B=5\times10^{-4}$, $\epsilon=0.1$, and $\eta=1$. Top panels: $T_i=0.2$. Bottom panels: $T_U=0.05$. Error bars represent SEM. Each point is averaged over 10,000  numerical simulations.
 }
 \label{fig4}
\end{figure}

The same analysis is done regarding the effect of local taxes on the opinion about the Union and the resulting possible withdrawals. Panels C-D of Fig. \ref{fig3} display the probability of experimenting withdrawal(s) from the Union in a period of 1000 steps versus the local tax rate $T_i$ for a fixed Union tax rate of $T_u=0.05$. Blue circles (red squares) correspond to a customs fee of $C_i=0.1$ ($0.2$). Panel C corresponds to the homogeneous scenario and Panel D to the heterogeneous one, with an unequal initial wealth distribution between and within countries. As shown, local taxes also positively correlate with exit event probability, although the trend is less pronounced than for Union taxes. The comparison between panels C and D confirm that, as shown before in the Union taxes analysis, heterogeneity in the wealth distribution increases withdrawals probability.


To check if these results are robust against other kinds of heterogeneities, we have also analyzed the case in which different countries present different conditions for business. Results can be found in Fig. \ref{figS1} 
of Appendix A, were the local profitabilities $\Gamma_{L,i}$ follow a normal distribution with mean $\mu=1.5$, and variance $\sigma=0.1$. Fig. \ref{figS1} 
confirms the previous trends: Union and local taxes positively correlate with exit(s) probability, although the Union integrity is more sensitive to Union taxes than to local ones. Also, by comparing Figs. \ref{fig3} and \ref{figS1} 
can be observed that heterogeneity in the business conditions promotes exits from the Union: the probability of exit events increases when some countries have better conditions for local business than others. Actually, heterogeneity in business conditions and unequal wealth distribution have similar effects on exits.

The role of countries' size heterogeneity on exits is studied in Figure \ref{figS2} 
of Appendix A, where countries' sizes $N_{i}$ are distributed according to $i\times 10^4;i=1,2,\ldots,20$. As shown by comparing Figs. \ref{fig3} and \ref{figS2},
 countries' size diversity does not have a significant effect on withdrawals.

\subsection*{Causes of withdrawals}

To identify the causes motivating exits from the Union, we analyze candidate factors, both at the country- and individual-level, at the moment of the withdrawal from the Union.

Regarding the effect of payoffs and taxes on exits, Fig. \ref{fig4} displays, at the moment of each withdrawal:
\begin{itemize}
    \item The relative difference between the \textit{per capita} wealth within the country leaving the Union and that in the rest of the Union (red squares).
    \item The \textit{per capita} wealth difference between anti-Union and pro-Union agents within the country leaving the Union (blue circles).    
    \item The \textit{per capita }wealth difference between anti-Union and pro-Union agents in the countries remaining in the Union (dark blue diamonds).
\end{itemize}
Panels A-B of Fig. \ref{fig4} show those observables versus the Union taxes rate. Panel A displays the results for the homogeneous scenario. Panel B shows the same results for the heterogeneous scenario. In the latter, the initial wealth distribution within each country follows a Poisson distribution with a mean taken, in turn, from a normal distribution ($\mu=10$,$\sigma^2=4$), resulting in wealth heterogeneity between and within countries.

Analyzing altogether these observables, the following conclusions hold:

\begin{itemize}
    \item For low Union taxes, countries with a higher risk to leave the Union are those in which anti-Union agents perform better than pro-Union agents. Nevertheless, there is no significant country performance difference between the leaving country and the rest of the countries in the Union.
    \item As Union taxes increase, these trends reverse. For high Union taxes, the countries that perform better show higher exit probability. Furthermore, there is no significant payoff difference between pro-Union and anti-Union agents in the leaving countries.
    \item For the countries that remain in the Union, no dependency of pro-Union and anti-Union agents’ payoff difference on the Union tax rate is observed. Note that, on average, within the Union, pro-Union agents perform better than anti-Union ones, being that difference independent of the Union tax burden.
\end{itemize}
To summarise, for low Union taxes the wealth difference within the country is the main exit promoter (imitation as a main driving force), while for high Union taxes the country performance is the main mechanism promoting exit (opinion dynamics as main driver).

Panels C-D of Fig. \ref{fig4} display the same results but as a function of local taxes instead of Union ones. As before, Panel C represents the homogeneous scenario and Panel D the heterogeneous scenario. From these results, one can conclude that:

\begin{itemize}
    \item Performance of leaving country is higher than the average in Union, being the difference independent of the local tax burden.
    \item For low local taxes, countries with a higher risk to leave the Union are those in which anti-Union agents perform better than pro-Union agents.
    \item As local taxes increase, the payoff differences between pro and anti-Union agents within countries decrease. It seems to be a direct consequence of the fact that local taxes redistribute wealth among the agents within the country.
\end{itemize}

In summarizing, for low local taxes, both payoffs difference within the country (imitation mechanism) and global country performance (opinion dynamics) coexist as exit drivers. For high local taxes, the opinion dynamics is the main mechanism driving exit.


By comparing left versus right panels of Fig. \ref{fig4}, it is shown that all these trends persist under heterogeneity in wealth distribution. As for the exit probabilities, we have checked the robustness for different heterogeneities by simulating the case in which different countries present different business conditions. In Fig. \ref{figS3} 
of Appendix A, countries' local profitability $\Gamma_{L,i}$ follows a normal distribution ($\mu=1.5$, $\sigma^2=0.15$). As it can be seen, all the previous trends remain. Furthermore, heterogeneity in business conditions displays similar results regarding payoff differences than unequal wealth distribution.


\section{Conclusions}
We have analyzed the dynamics of an economic Union through an agent-based model that replicates a collection of countries, which can or cannot be part of the Union, composed of agents that interact among them through both business and opinion exchange. The main goal of the study is to characterize a set of plausible factors that could motivate withdrawals from the Union.

Through numerical simulations, we have shown that taxes promote the risk of withdrawal, while high customs across the boundaries of the Union promote Union integrity. Interestingly, we find that the effect of Union taxes in raising the risk of withdrawal is stronger than that of local taxes. Regarding the drivers for Union exists, we have identified that, for low Union taxes, the wealth difference within the country is the leading cause of withdrawals; as Union taxes increase, the global country's performance turns to be the determining cause for leaving the Union. Furthermore, while for low local taxes, wealth difference and country performance are withdrawal causes, as local taxes increase, the differences between countries' wealth constitute the dominant cause driving exit. We have also shown that inequality, either in wealth or business conditions, promotes withdrawals. Conversely, unequal country sizes do not affect exits. As a prospective remark, it will be of interest to study in detail the effect of some other factors such as the costs associated with maintaining the Union and the opinions about immigration and shared policies. To conclude, we hope the findings here will be of interest to policy-makers when designing economic and taxation strategies and campaigns to inform citizens about unions. 

\section*{Acknowledgments}

C.G.L. and Y.M. acknowledge partial support from the Government of Aragon and FEDER funds, Spain through grant E36-20R (FENOL), by MCIN/AEI and FEDER funds (grant PID2020-115800GB-I00), and from Banco Santander (Santander-UZ 2020/0274). The funders had no role in study design, data collection, and analysis, decision to publish, or preparation of the manuscript.
\nolinenumbers

\bibliography{brexit.bib}

\newpage
\section*{Appendices}
\subsection*{Appendix A: Supplementary figures}
\label{Supp.Figures}

\begin{figure}[ht!]
\centering
\includegraphics[width=\columnwidth]{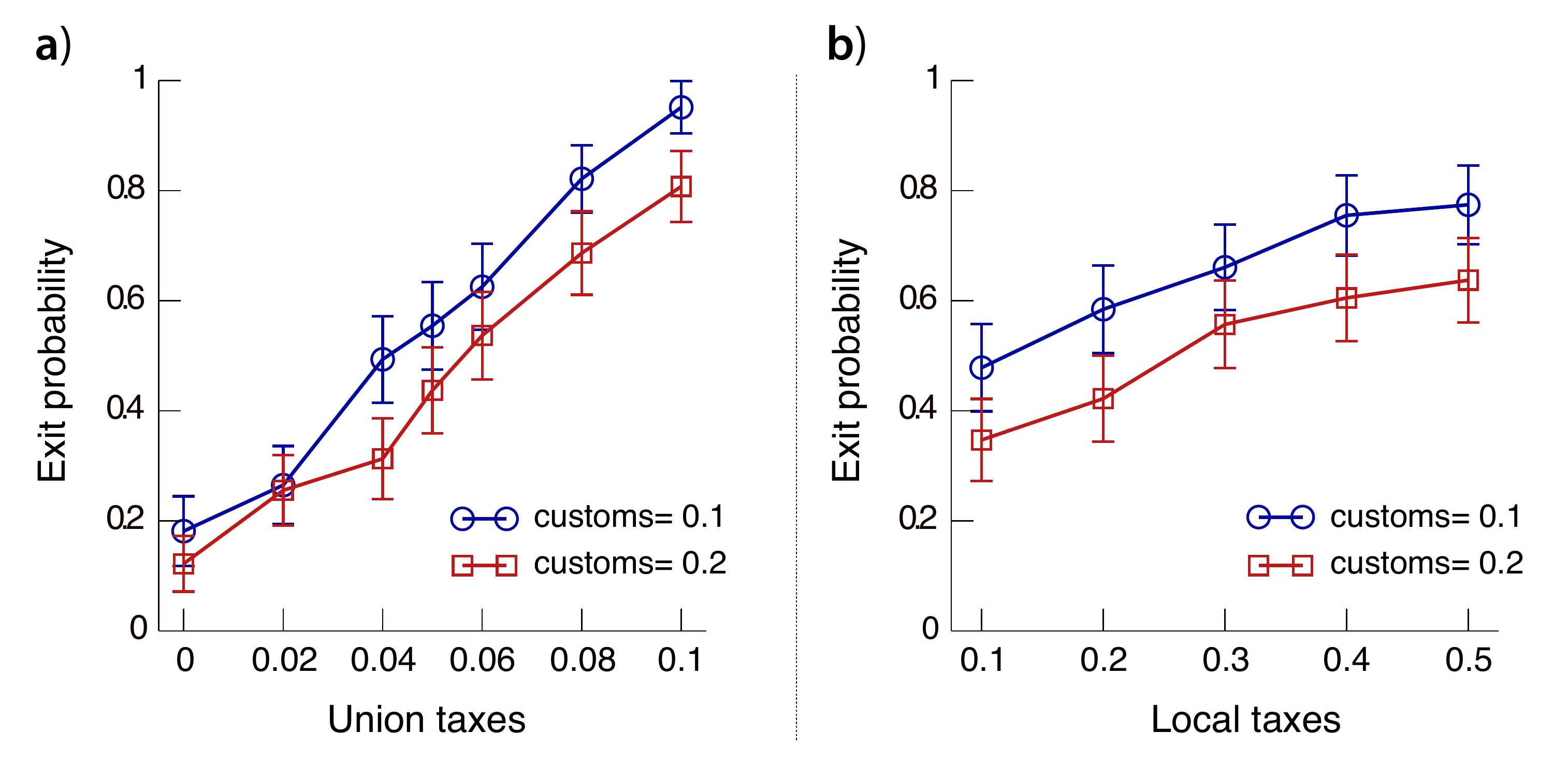} 
\caption{\textbf{Withdrawal probability under unequal business' conditions}. Probability of at least an exit event (i.e., a country leaves the Union) in a finite interval of 1,000 rounds ($\sim20$ years) as a function of the Union tax rate (panel \textbf{A}) and local tax rate (\textbf{B}). Local tax rate was fixed to $T_i=0.2$ in Panel \textbf{A}, and Union tax rate to $T_U=0.05$ in Panel \textbf{B}. Each point is averaged over 10,000  numerical simulations. In these plots, different countries provide different conditions for local businesses: local profitabilities $\Gamma_{L,i}$ follow a normal distribution with $\mu=1.5$, $\sigma^2=0.15$. $C=20$, $N_i=10^5$, $\Gamma_G=\Gamma_U=1.5$, $\psi=0.8$, $\rho=0.95$, $x=20$, $B=5\times10^{-4}$, $\epsilon=0.1$, $\eta=1$. See further details in the main text.}
 \label{figS1}
\end{figure}
\begin{figure}[ht!]
\centering
 \includegraphics[width=\columnwidth]{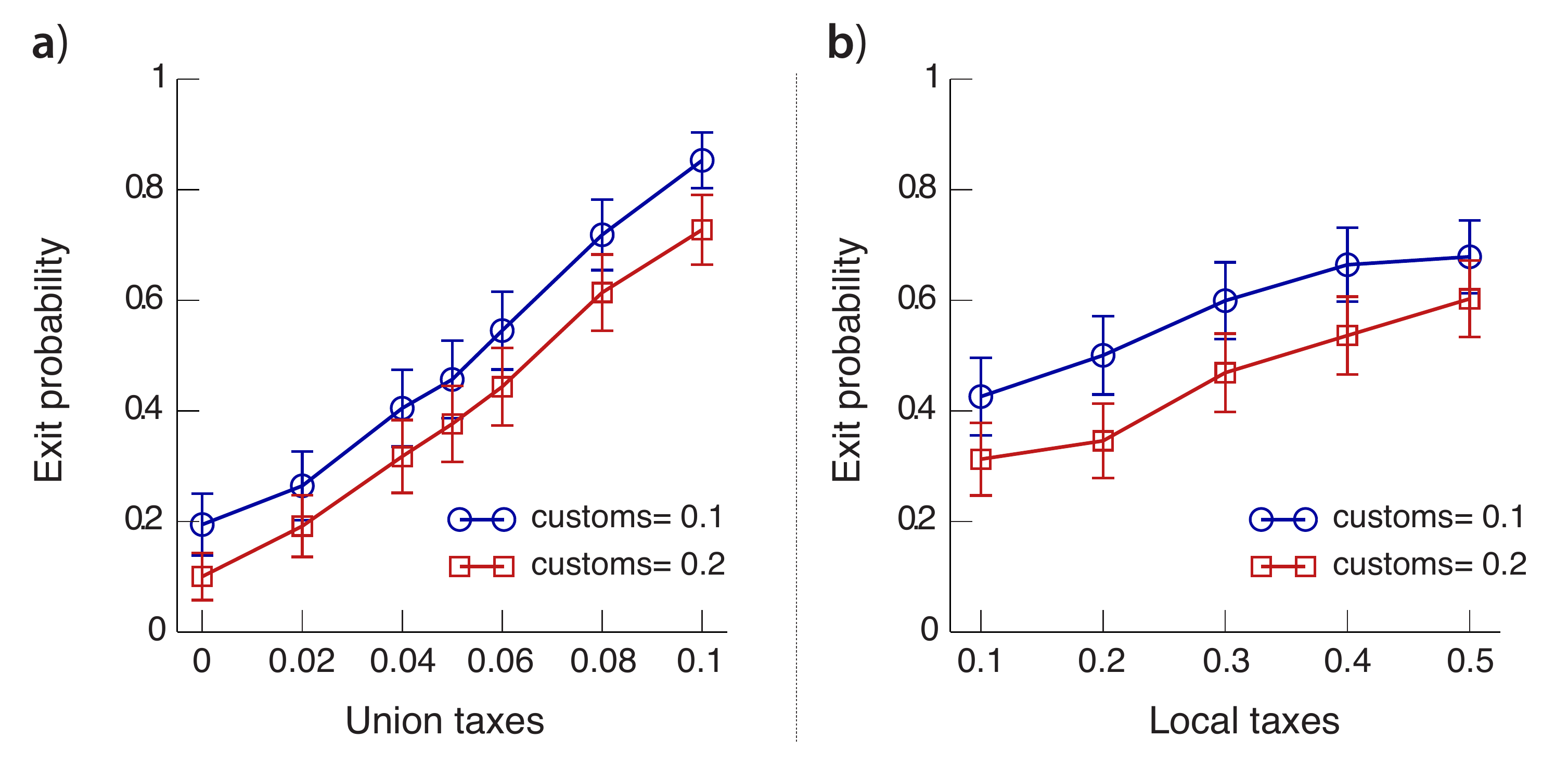} 
 \caption{\textbf{Withdrawal probability under unequal countries' sizes}. Probability of at least a withdrawal in a finite interval of 1,000 ($\sim20$ years) rounds as a function of the Union tax rate (panel \textbf{A}) and local tax rate (\textbf{B}). In Panel \textbf{A}, local taxes are fixed to $T_i=0.2$. In Panel \textbf{B}, Union taxes are fixed to $T_U=0.05$. In this plots, countries' sizes $N_{i}$ are distributed according to $i\times 10^4;i=1,2,\ldots,20$. Each point is averaged over 10,000  numerical simulations. $C=20$, $\Gamma_i=\Gamma_G=\Gamma_U=1.5$, $\psi=0.8$, $\rho=0.95$, $x=20$, $B=5\times10^{-4}$, $\epsilon=0.1$, $\eta=1$. See further details in the main text.}
 \label{figS2}
\end{figure}
\begin{figure}[ht!]
\centering
 \includegraphics[width=\columnwidth]{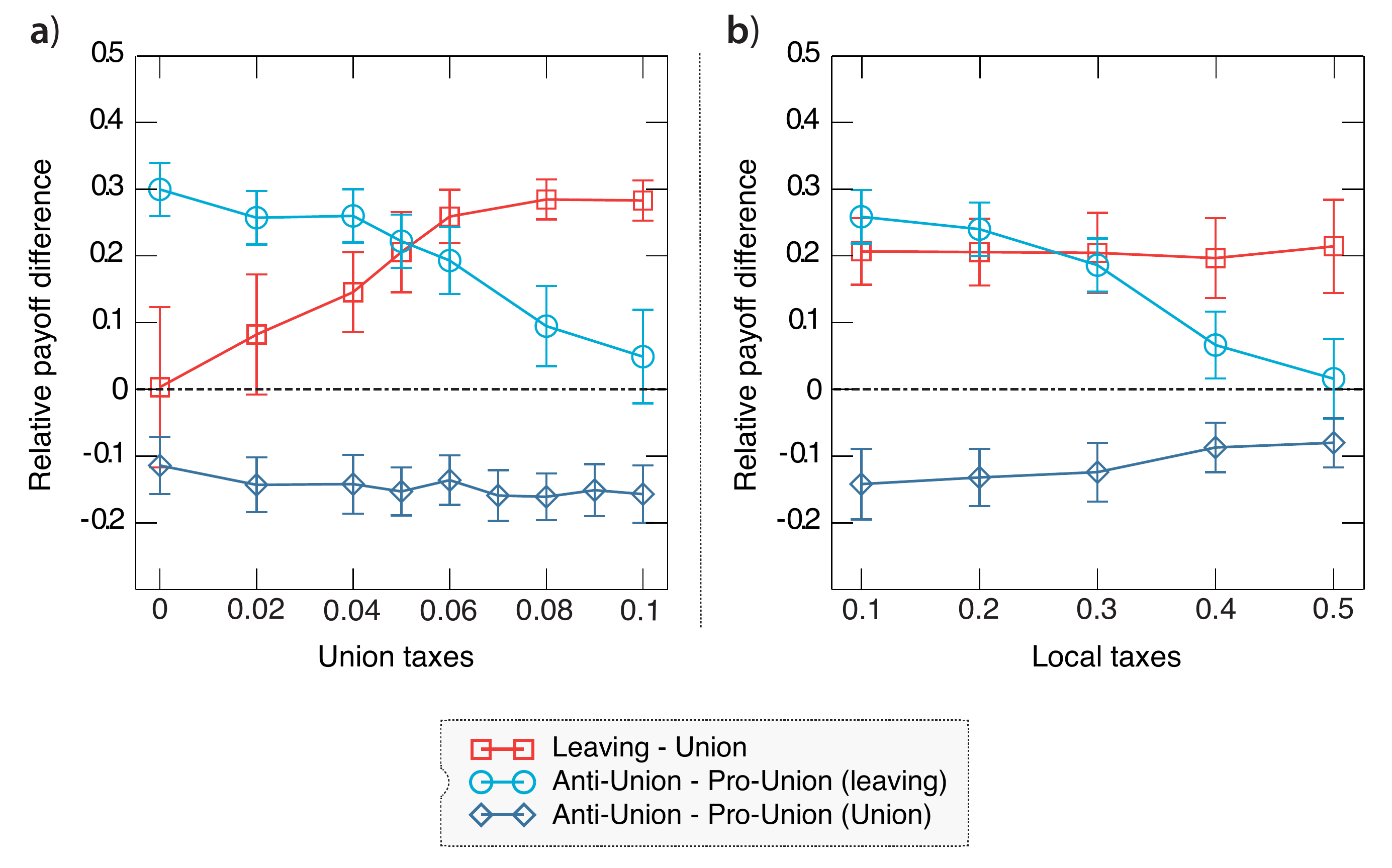} 
 \caption{\textbf{Exit drivings under unequal business' conditions.} Payoff differences at the moment of an exit versus Union tax rate (panel \textbf{A}) and local tax rate (\textbf{B}). Red squares represent the relative difference between the accumulated payoff in the country that leaves the Union and the average accumulated payoff in the Union at the moment of an exit. Green circles (resp., blue diamonds) represent, at the moment of an exit, the difference between the average accumulated payoff of anti-Union agents and pro-Union agents in the country leaving the Union (resp., in the Union). 
 Here, as in Fig. \ref{figS1}, local business' profitabilities $\Gamma_{L,i}$ follow a normal distribution with $\mu=1.5$, $\sigma^2=0.15$. $C=20$, $N_i=10^5$, $\Gamma_G=\Gamma_U=1.5$, $\psi=0.8$, $\rho=0.95$, $x=20$, $B=5\times10^{-4}$, $\epsilon=0.1$, $\eta=1$. See further details in the main text.}
 \label{figS3}
\end{figure}

In this appendix, we display complementary figures to those in the main text. Figs. \ref{figS1} and \ref{figS2} complement the results shown in Fig. \ref{fig3} of the main text; Fig. \ref{figS3} complements Fig. \ref{fig4}.

Figure \ref{figS1} shows the withdrawal probability for the case in which the conditions for business depend on the country. Explicitly, local profitabilities $\Gamma_{L,i}$ follow a normal distribution ($\mu = 1.5$, $\sigma = 0.1$).  As in the case of a homogeneous distribution (Fig. \ref{fig3}),  withdrawal probability increases with Union and local taxes. Furthermore, by comparing both figures,
it is shown that withdrawal probability increases with business conditions heterogeneity.

Fig. \ref{figS2} displays the withdrawal probabilities for unequal countries’ sizes. Explicitly, sizes are distributed according to $N_i$=10,000, 20,000, 30,000, ..., 200,000. The comparison of Fig. S1 and Fig. 3 of the main text shows that the effect of countries' sizes diversity on withdrawal probability is not significant.

To complement Fig. \ref{fig4}, Fig. \ref{figS3} shows the wealth differences at the moment of a withdrawal for heterogeneity in business conditions. The comparison of both figures shows that dependencies are robust against business conditions distribution.

\newpage

\end{document}